\documentclass[11pt,a4paper]{article}
\usepackage[body={16.5cm,25cm}]{geometry}
\usepackage[dvips]{color}
\usepackage{array}
\usepackage{stackrel}
\usepackage{tensor}
\usepackage{mathtools}
\usepackage{curves}
\usepackage[titletoc]{appendix}

\usepackage[hang,small,center]{caption}
\usepackage{amsmath,amssymb}
\usepackage{amsfonts}
\usepackage{mathrsfs}
\usepackage{epic}
\usepackage{eepic}
\usepackage{graphicx}
\usepackage{cite}
\newcommand{\ds}{\displaystyle}

\renewcommand{\author}[1]{\large\rm #1\\ \bigskip}
\newcommand{\address}[1]{{\normalsize\it #1\\}\bigskip}
\renewcommand{\title}[1]{\bigskip\bigskip\Large\bf #1\bigskip\bigskip\\}

\newcommand{\Bigpsi}[3]{\phantom{\Psi}_2 \kern -.05em
\Psi_2\left(\genfrac{}{}{0pt}{}{#1}{#2}\biggl|#3\right)}

\newmuskip\pfqmuskip
\newcommand*\pfq[6][8]{%
  \begingroup 
  \pfqmuskip=#1mu\relax
  \mathchardef\normalcomma=\mathcode`,
  \mathcode`\,=\string"8000
  \begingroup\lccode`\~=`\,
  \lowercase{\endgroup\let~}\pfqcomma
  {}_{#2}\phi_{#3}{\left[\genfrac..{0pt}{}{#4}{#5};#6\right]}%
  \endgroup
}
\newcommand{\pfqcomma}{{\normalcomma}\mskip\pfqmuskip}

\DeclareMathOperator{\Tr}{Tr}

\def\bea{\begin{eqnarray}}
\def\eea{\end{eqnarray}}
\newcommand{\be}{\begin{equation}}
\newcommand{\ee}{\end{equation}}
\newcommand{\beq}{\begin{equation}}
\newcommand{\eeq}{\end{equation}}

\newcommand{\ga}{{\gamma}}
\newcommand{\al}{{\alpha}}
\newcommand{\de}{{\delta}}
\newcommand{\bet}{{\beta}}
\newcommand{\ep}{{\epsilon}}

\newcommand{\overbar}[1]{\mkern 1.5mu\overline{\mkern-1.5mu#1\mkern-1.5mu}\mkern 1.5mu}
\def\mathLarge#1{\mbox{\large $#1$}}

\newcommand{\lam}{{\lambda}}

\newcommand{\phiW}{\phantom{|}_8{{W}}{_7}}

\newcounter{app}
\newcounter{sapp}[app]

\def\nsection#1{\setcounter{equation}{0}\section{#1}}

\begin{document}

\vglue 2 cm
\begin{center}
\title{Boundary matrices for the higher spin six vertex model}
\author{Vladimir
  V.~Mangazeev$^{1}$ and Xilin Lu$^{1}$}
\address{$^1$Department of Theoretical Physics,
         Research School of Physics and Engineering,\\
    Australian National University, Canberra, ACT 0200, Australia.
}

\begin{abstract}
In this paper we consider solutions to the reflection equation related
to the higher spin stochastic six vertex model. The corresponding
higher spin $R$-matrix is associated with the affine quantum algebra
$U_q(\widehat{sl(2)})$.
The explicit formulas for  boundary $K$-matrices for spins $s=1/2,1$  are well known.
We derive difference equations for the generating function of  matrix elements
of the $K$-matrix for any spin $s$
and solve them in terms of hypergeometric functions.
As a result we derive the explicit formula for matrix elements of the $K$-matrix
for arbitrary spin. In the lower- and upper- triangular cases, the $K$-matrix
simplifies and
reduces to simple products of $q$-Pochhammer symbols.

\end{abstract}

\end{center}

\newpage

\nsection{Introduction}

In the last two decades years there was a significant growth of interest in applications of
quantum integrable systems to KPZ universality \cite{KPZ86}, stochastic processes and
non-equilibrium statistical mechanics \cite{Der98,Cor12,BP16}.
The asymmetric simple exclusion process (ASEP) \cite{SP70} is one of the most studied examples,
both on a line and with open boundary conditions (see, for example, \cite{DEHP93,Sas99,DGE05,TW09}).
It is intimately connected to the higher spin stochastic six vertex model which has been
studied on a quadrant or a semi-infinite line
with simple open boundary conditions
\cite{BorCG2016,BBCW2018,BP2018}. The $R$-matrix of the higher spin six vertex model
is related to the higher weight representations
of the  $U_q(\widehat{sl(2)})$ algebra and the explicit formula was derived in \cite{Man14}.

One of the main approaches to quantum integrable systems
with general open boundary conditions is Sklyanin's method \cite{Skl88}. This method relies on solutions
of the reflection equation  \cite{Cherednik:1985vs,Skl88}.
In principle, solutions of the reflection
equation for higher spins can be obtained using the fusion procedure \cite{Nepom92,Nepom07}
but such formulas are not explicit
and quite complicated.

In this paper we attempt to find a general explicit expression for reflection matrices for
the higher spin
six vertex model in a stochastic gauge. Starting with the
four-parametric solution of the reflection equation for the spin $s=1/2$ \cite{vega1993}, we get explicit
formulas for matrix elements of the $K$-matrix for any higher spin.

The paper is organized as follows. In Section 2 we review the  theory of reflection equations
and the construction of commuting transfer-matrices with open boundary conditions. We also
slightly generalize it to include models with $R$-matrices lacking a difference property.
In Section 3 we review the construction of the $R$-matrix for the higher spin six vertex model
and its factorization properties. In Section 4 we derive the recurrence relations for matrix elements
of the higher spin $K$-matrices. In Section 5 we solve these recurrence relations in special
low- and upper- triangular cases. In Section 6 we introduce equations for the generating function of the
matrix elements of the $K$-matrix in a non-degenerate case. In Section 7 we solve these equations
and find a solution for the generating function in terms of the terminating balanced
${}_4\phi_3$ series. We also obtain the explicit formula for matrix elements in the form
of a double sum. Finally, in Conclusion we discuss the obtained results and outline directions for further
research.

\section{Reflection equation and commuting transfer-matrices}

Reflection equation \cite{Cherednik:1985vs,Skl88} plays a fundamental role in constructing quantum integrable systems
with open boundary conditions. For a given solution $R_{12}(x,y)$
of the Yang-Baxter equation
\beq
R_{12}(x,y)R_{13}(x,z)R_{23}(y,z)=R_{23}(y,z)R_{13}(x,z)R_{12}(x,y), \label{intr1}
\eeq
the reflection equation has the following form
\begin{equation}
R_{12}(x,y)K_1(x)R_{21}(y,\bar{x})K_2(y)=K_2(y)R_{12}(x,\bar{y})K_1(x)R_{21}(\bar{y},\bar{x}).
\label{intr2}
\end{equation}
Here we assume that the $R$-matrix $R_{12}(x,y)$ is a linear operator acting nontrivially
in the tensor product of vector spaces $V_1\otimes V_2$ and $K_i(x)$  acts nontrivially in $V_i$, $i=1,2$.
In general, the $R$-matrix $R_{12}(x,y)$ does not have a difference property
and the variables $\bar{x}$, $\bar{y}$ are the ``reflected'' spectral parameters.
For trigonometric $R$-matrices we have
\beq
R_{12}(x,y)=R_{12}(x/y) \label{intr2c}
\eeq
 and $\bar{x}=x^{-1}$.

If the $R$-matrix is regular, i.e.
\beq
R_{12}(x,x)=\mathcal{P}_{12} \label{intr2a}
\eeq
 with
$\mathcal{P}_{12}$ being a permutation operator, then
$R_{12}(x,y)$ also satisfies the unitarity condition
\begin{equation}\label{intr2b}
R_{12}(x,y)R_{21}(y,x)=f(x,y){{I\otimes I}}.
\eeq

Using the $R$-matrix  $R_{12}(x,y)$ and the boundary matrix $K(x)$ we can construct
a double row monodromy matrix acting in the tensor product $V_1\otimes\ldots\otimes V_L$
\beq
T_a(x)=R_{a1}({x},z_1)\cdots R_{aL}({x},z_L)K_a({x})R_{La}(z_L,\bar{x})\cdots R_{1a}(z_1,\bar{x}).\label{intr3}
\eeq
with $V_a$ being the auxiliary space.

It is easy to show that as a consequence of (\ref{intr1}-\ref{intr2}) the double row monodromy
matrix
satisfies the relation
\begin{equation}\label{intr4}
R_{ab}(x,y)T_a(x)R_{ba}(y,\bar{x})T_b(y)=T_b(y)R_{ab}(x,\bar{y})T_a(x)R_{ba}(\bar{y},\bar{x}).
\end{equation}

Let us assume that $R_{12}^{t_2}(x,y)$ is non-degenerate and
define a linear operator \cite{Dancer}
\beq
\mathcal{R}_{12}(x,y)=[(R_{21}(y,x)^{t_1})^{-1}]^{t_1}.\label{intr5}
\eeq
which implies
\beq
\mathcal{R}_{12}^{t_1}(x,y)R_{21}^{t_1}(y,x)=\mathcal{R}_{12}^{t_2}(x,y)R_{21}^{t_2}(y,x)={{I\otimes I}}.
\label{intr5a}
\eeq

We define the dual reflection equation  by
\begin{equation}\label{intr6}
R_{12}(\bar{x},\bar{y})\bar{K}_1(x)\mathcal{R}_{21}(\bar{y},x)\bar{K}_2(y)=
\bar{K}_2(y)\mathcal{R}_{12}(\bar{x},y)\bar{K}_1(x)R_{21}(y,x),
\end{equation}

For a given solution $\bar{K}(x)$ of \eqref{intr6} and the monodromy matrix \eqref{intr3}
 we define a  double row transfer matrix $t(x)$ as
\begin{equation}
t(x)=\Tr_a\{\bar{K}_a(x)T_a(x)\},\label{intr7}
\end{equation}
Then double row transfer matrices \eqref{intr7} commute
\beq
[t(x),t(y)]=0.\label{intr8}
\eeq
We notice that we do not require the crossing unitarity of the $R$-matrix,
only the existence of $\mathcal{R}_{12}(x,y)$ and $\bar{K}(x)$.

The proof goes as follows
\begin{equation}\label{intr9}
\begin{aligned}
&\phantom{=f}f(x,y)t(x)t(y)\\
&= \Tr_{a,b}\{ f(x,y)\bar{K}_{a}(x)T_{a}(x)\bar{K}_{b}(y)T_{b}(y) \}
=\Tr_{a,b}\{ f(x,y)\bar{K}_{b}(y)\bar{K}_{a}^{t_a}(x)T_{a}^{t_a}(x)T_{b}(y) \}\\
&=\Tr_{a,b}\{ f(x,y)\bar{K}_{b}(y)\bar{K}_{a}^{t_a}(x)[\mathcal{R}_{ab}^{t_a}
(\bar{x},y)R_{ba}^{t_a}(y,\bar{x})] T_{a}^{t_a}(x)T_{b}(y) \}\\
&=\Tr_{a,b}\{ f(x,y)\bar{K}_{b}(y)[\mathcal{R}_{ab}(\bar{x},y)\bar{K}_{a}(x)]^{t_a}
[T_{a}(x)R_{ba}(y,\bar{x})]^{t_a} T_{b}(y) \}\\
&=\Tr_{a,b}\{[\bar{K}_{b}(y)\mathcal{R}_{ab}(\bar{x},y)\bar{K}_{a}(x)R_{ba}(y,x)]
[R_{ab}(x,y)T_{a}(x)R_{ba}(y,\bar{x}) T_{b}(y) ]\},
\end{aligned}
\end{equation}
where we used \eqref{intr2b}, \eqref{intr5a} and the fact that $\Tr(AB)=\Tr(A^tB^t)$
for any matrix operators $A$ and $B$.

Now using \eqref{intr4} and \eqref{intr6} we transform \eqref{intr9} to
\begin{equation}\label{intr10}
\begin{aligned}
&\phantom{=}\Tr_{a,b}\{[R_{ab}(\bar{x},\bar{y})\bar{K}_{a}(x)\mathcal{R}_{ba}(\bar{y},x)\bar{K}_{b}(y)]
[T_{b}(y)R_{ab}(x,\bar{y})T_{a}(x)R_{ba}(\bar{y},\bar{x})] \}\\
&=\Tr_{a,b}\{R_{ba}(\bar{y},\bar{x})R_{ab}(\bar{x},\bar{y})
\bar{K}_{a}(x)\bar{K}_{b}^{t_b}(y)\mathcal{R}_{ba}^{t_b}(\bar{y},x)R_{ab}^{t_b}(x,\bar{y})T_{b}^{t_b}(y)T_{a}(x) \}\\
&=\Tr_{a,b}\{f(\bar{x},\bar{y})\bar{K}_{a}(x)\bar{K}_{b}^{t_b}(y)T_{b}^{t_b}(y)T_{a}(x) \}
=\Tr_{a,b}\{ f(x,y)\bar{K}_{b}(y)T_{b}(y)\bar{K}_{a}(x)T_{a}(x) \}\\
&=f(x,y)t(y)t(x),
\end{aligned}
\end{equation}
i.e. we showed a commutativity of two transfer-matrices \eqref{intr8}.

If the $R$-matrix satisfies the difference property \eqref{intr2c} and the crossing
unitarity condition
\beq
(M_1^{-1}R_{12}(\lam)M_1)^{t_1}R_{21}^{t_1}((\rho^2\lam)^{-1})=g(\lambda){{I\otimes I}} \label{intr11}
\eeq
for some $\rho\in\mathbb{C}$ and a constant matrix $M\in \mbox{End}(V)$, then
$\mathcal{R}_{12}(x,y)=\mathcal{R}_{12}(x/y)$ exists and is given by
\beq
\mathcal{R}_{12}(\lam)=\frac{1}{g(\lambda/\rho^2)}M_1^{-1}R_{12}(\lambda/\rho^2)M_1.\label{intr12}
\eeq
In general, \eqref{intr11} is a stronger condition than \eqref{intr5} even for $R$-matrices
with a difference property \cite{Dancer}.

Using \eqref{intr12} we can map the dual reflection equation
\eqref{intr6} to \eqref{intr2}. If the matrix $M$ satisfies the property
\beq
[M\otimes M, R_{12}(\lam)]=0, \label{spec40}
\eeq
then we have a solution to \eqref{intr6}
\beq
\bar{K}(x)=M^{-1}\,K({1}/({qx})).\label{spec4}
\eeq
Notice that \eqref{spec4} can be used to construct solutions of \eqref{intr6} from any solution
$K(x)$ of \eqref{intr2} provided that $\mathcal{R}_{12}(\lam)$ is given by \eqref{intr12}.
There are other automorphisms between solutions of the reflection equation and its dual \cite{Skl88} but
we will not consider them here.

\nsection{The higher spin six vertex model}

In this section we start with explicit formulas for the higher-spin $R$-matrix $R_{I,J}(\lam)$
related to the $U_q(\widehat{sl(2)})$ algebra following  \cite{Man14,BM2016}.

For arbitrary complex weights $I,J\in\mathbb{C}$ we define a linear operator
$R_{I,J}(\lam)\in\mbox{End}(V\otimes V)$
by its action on the basis  $|i\rangle$, $i\in\mathbb{Z}_+$ of $V$
\beq
R_{I,J}(\lam)|i'\rangle\otimes|j'\rangle=\sum_{i,j}R_{I,J}(\lambda)_{i,j}^{i',j'}
|i\rangle\otimes|j\rangle,\label{high1}
\eeq
where matrix elements $R_{I,J}(\lambda)_{i,j}^{i',j'}$ are given by the following expression \cite{BM2016}
\beq
\begin{aligned}
R_{I,J}(\lambda)_{i,j}^{i',j'}=& \delta_{i+j,i'+j'} q^{i'j'-ij-iJ-Ij'}\begin{bmatrix}
i+j\\i
\end{bmatrix}_{q^2}\frac{(\lambda^{-2}q^{I-J};q^2)_{j'}
(\lambda^{-2}q^{J-I};q^2)_{i}(q^{-2J};q^2)_j}{(\lambda^{-2}q^{-I-J};q^2)_{i+j}(q^{-2J};q^2)_{j'}}\\
&\times \pfq{4}{3}{q^{-2i},q^{-2j'},\lambda^2 q^{-I-J},\lambda^2q^{2+I+J-2i-2j}}
{q^{-2i-2j},\lambda^2q^{2+I-J-2i},\lambda^2q^{2+J-I-2j'}}{q^2,q^2}.
\end{aligned}\label{Rmatrix}
\eeq
Here we used standard notations for $q$-Pochhammer symbol, $q$-binomial coefficients and the basic hypergeometric
function $\phantom{|}_4\phi_3$ (see Appendix A).

The $R$-matrix \eqref{Rmatrix} satisfies the Yang-Baxter equation \eqref{intr1} with three arbitrary
weights $I,J,K\in\mathbb{C}$ associated with $V_{1,2,3}$. Let us notice that an apparent singularity
coming from $q^{-2i-2j}$ for $i,j\in\mathbb{Z}_+$ in \eqref{Rmatrix} never happens, since
the sum terminates earlier either at $i\leq i+j$ or $j'\leq i+j$ due to a conservation law $i+j=i'+j'$.
Therefore, the hypergeometric function in \eqref{Rmatrix}  does not require a regularization.
The representation \eqref{Rmatrix} is equivalent to (5.8) from \cite{Man14} after a Sears'
transformation \eqref{Sears}.

From now on we will assume that weights $I$ and $J$ are positive integers and
the $R$-matrix acts in the tensor product $V_I\otimes V_J$, where $V_I$ is a finite-dimensional
module with the basis $|i\rangle$, $i=0,\ldots,I$. Therefore, we will be looking at finite-dimensional
solutions of the Yang-Baxter and reflection equation unless explicitly stated otherwise.

The reason for this is that the Sklyanin approach to integrable systems  with open boundaries \cite{Skl88}
relies on the crossing relation. As shown in the previous section
this can be relaxed to the existence of the operator
$\mathcal{R}_{12}$ in \eqref{intr5}. A sufficient condition for the operator $\mathcal{R}_{12}$
to exist is the crossing symmetry of the $R$-matrix (or a weaker condition of the crossing unitarity).

To our knowledge the crossing symmetry for the $R$-matrix \eqref{Rmatrix}
is only known when $I,J\in\mathbb{Z}_+$. To write it down it is convenient to define
a symmetric version of \eqref{Rmatrix}
\beq
\bar{R}_{I,J}(\lam)_{i,j}^{i',j'}=\lam^{i-i'}R_{I,J}(\lam)_{i,j}^{i',j'} \label{high0}.
\eeq
In particular, $\bar{R}_{I,J}(\lam)$ with $I=J=1$ is proportional to the $R$-matrix of the symmetric 6-vertex model.

Let us use the standard notation $\bar{R}_{12}(\lam)$ for \eqref{high0} and assume that the first and the second spaces
correspond to representations with weights $I$ and $J$, respectively.

Then we have two relations
\beq
\mathcal{P}_{12} \bar{R}_{12}(\lam)\mathcal{P}_{12}=\bar{R}_{12}(\lam),\quad I,J\in\mathbb{Z}_+, \label{high1c}
\eeq
and
\beq
f_{I,J}(\lam)\bar{R}_{12}(\lam)_{ij}^{i',j'}=\frac{c_{i,I}}{c_{i',I}}\bar{R}_{12}
((q\lam)^{-1})_{I-i',j}^{I-i,j'}\label{high1a}
\eeq
with
\beq
f_{I,J}(\lam)=q^{-IJ}\frac{(\lam^2q^{2-I+J};q^2)_I}{(\lam^2q^{2-I-J};q^2)_I},
\quad c_{i,I}=q^{i(i+1)}\frac{(q^{-2I};q^2)_i}{(q^2;q^2)_i}. \label{high1b}
\eeq
The relation \eqref{high1c} was proved in \cite{Man14}. The second relation
\eqref{high1a} with $I\in\mathbb{Z}_+$ can be proved  by using Sears' transformation \eqref{Sears}.
It is done in three steps. First we apply to the RHS of \eqref{high1a} the transformation
 \eqref{Sears} with $q\to q^2$ and the following choice of parameters $n=j'$, $a=q^{-2(I-i')}$,
 $b=\lam^{-2}q^{-2-I-J}$, $c=\lam^{-2}q^{J-I+2i'-2j}$, $d=\lam^{-2}q^{J-I-2j'}$,
 $e=q^{-2(I-i'+j)}$ and $f=\lam^{-2}q^{-I-J+2i'}$. Second, we use relation \eqref{high1c}
 to interchange $I$ and $J$ and all indices between the first and the second spaces. The result
 coincides with (5.8) from \cite{Man14} up to a certain factor. Applying Sears' transformation again
we come to the LHS of \eqref{high1a}.

We can rewrite the relation \eqref{high1a} as a  crossing relation
\beq
f_{I,J}(\lam)\bar{R}_{12}(\lam) =V_1 \bar{R}_{21}^{t_1}((q\lam)^{-1})\>V_1^{-1}, \label{high1d}
\eeq
where $V_1$ is a $(I+1)\times(I+1)$ matrix with matrix elements
\beq
V_{i,j}=c_{i,I}\delta_{I-i,j},\quad i,j=0,\ldots,I.\label{high1e0}
\eeq
As an immediate consequence of \eqref{high1d} and the inversion relation
\beq
\bar R_{12}(\lam)\bar R_{21}(\lam^{-1})={I\otimes I}\label{high1f}
\eeq
 we have a crossing unitarity relation
\beq
\bar{R}_{12}^{t_1}(\lam)\bar{R}_{21}^{t_1}(1/(q^2\lam))=g_{IJ}(\lambda)\>{I\otimes I}, \label{high1e}
\eeq
where
\beq
g_{IJ}(\lam)=\frac{f_{IJ}(q\lam)}{f_{IJ}(\lam)}=\frac{(1-\lam^2q^{2+I+J})(1-\lam^2q^{2-I-J})}
{(1-\lam^2q^{2+I-J})(1-\lam^2q^{2-I+J})}.\label{high1g}
\eeq
The easiest way to see that the inversion factor in \eqref{high1f} is equal to $1$ is
to rewrite it in terms of the stochastic $R$-matrix \eqref{Sdef} and use the relation \eqref{high2a} below.

Following \cite{KMMO16,BM2016} we introduce a
stochastic version of the higher-spin
six-vertex model  with the $R$-matrix
\begin{equation}\label{Sdef}
S_{I,J}(\lambda)_{i,j}^{i',j'}=q^{ij-i'j'-Ji+Ij'}R_{I,J}(\lambda)_{i,j}^{i',j'}.
\end{equation}
Using the conservation laws $i+j=i'+j'$ for all $R$-matrices in \eqref{intr1}
one can easily show that the twist in \eqref{Sdef} does not affect the Yang-Baxter equation

Let us  define the following function
\begin{equation}\label{high3}
\Phi_q (\gamma|\beta;x,y)= \left(\frac{y}{x}\right)^{\gamma}\frac{(x ;q)_{\gamma}
\left(y/x;q\right)_{\beta-\gamma}}{(y;q)_{\beta}} \begin{bmatrix}
\beta \\ \gamma
\end{bmatrix}_q.
\end{equation}
This function was introduced in \cite{KMMO16} for an arbitrary rank $n$ of the $U_q(A_{n}^{(1)})$ algebra.
Here we only consider the case $n=1$.

The stochastic R-matrix \eqref{Sdef} admits the following factorization \cite{BM2016} in terms of
$\Phi$ functions:
\begin{equation}\label{Sfactor}
S_{I,J}(\lambda)_{i,j}^{i',j'}=\delta_{i+j,i'+j'}\sum_{m+n=i+j}
\Phi_{q^2}\left(m-j|m;\frac{q^{J-I}}{\lambda^2},\frac{q^{-I-J}}
{\lambda^2}\right)\Phi_{q^2}\left(n|j';\frac{\lambda^2}{q^{I+J}},q^{-2J}\right).
\end{equation}

The $R$-matrix \eqref{Sfactor} satisfies the stochasticity condition \cite{KMMO16,BM2016}
\begin{equation}
\sum_{i,j}S_{I,J}(\lambda)_{i,j}^{i',j'}=1.\label{high2}
\end{equation}

The proof  immediately follows from the identity
\beq
\sum_{0\leq\gamma\leq\beta}\Phi_q(\gamma|\beta;x,y)=1,\label{sumrule}
\eeq
which we apply twice to \eqref{Sfactor}.

Let us notice that the inversion relation for $S$
\beq
S_{12}(\lam)S_{21}(\lam^{-1})={I\otimes I} \label{high2a}
\eeq
follows from the Yang-Baxter equation and \eqref{high2}.

It is easy to see that the crossing unitarity relation \eqref{high1e}
for the stochastic $R$-matrix $S_{12}(\lam)$ takes the following form
\beq
M_1S_{12}^{t_1}(\lam)M_1^{-1}S_{21}^{t_1}((q^2\lam)^{-1})=g_{IJ}(\lambda)\>{I}\otimes{ I}, \label{high4}
\eeq
where
\beq
M=\mbox{diag}(1,q^2,\ldots,q^{2I}) .\label{high5}
\eeq

One can ask  whether the relation \eqref{high4} can be generalized to arbitrary
$I,J\in\mathbb{C}$, since the $R$-matrix  \eqref{Sfactor}
is well defined in this case \cite{KMMO16}.
The answer is apparently negative.
If we substitute \eqref{Sfactor} directly into \eqref{high4}, we get a triple sum,
with two summations coming from \eqref{Sfactor} and a single sum coming from the summation over
matrix indices in \eqref{high4}. After straightforward calculations
one can see that this last sum is given again by a balanced $\phantom{|}_4\phi_3$ series
which terminates when either $I$ or $J$ is a positive integer. Then we can use Sears' transformations
to prove \eqref{high4} directly. When both $I,J\in\mathbb{C}$, no transformation
between two non-terminating  $\phantom{|}_4\phi_3$ series exists. A simple numerical check
shows that \eqref{high4} does not hold in this case.
However, the operator \eqref{intr5} may still
exist and can be used to define the dual reflection equation.

Finally we notice
that there are several choices of the spectral parameter $\lam$, when the $R$-matrix $S_{I,J}(\lam)$
simplifies to a factorized form. First, it is easy to check two properties of the function $\Phi_q$
\begin{equation}
\Phi_q(i|j;1,y)=\delta_{i,0},\quad \Phi_q(i|j;y;y)=\delta_{i,j}. \label{high7}
\end{equation}
Substituting $\lam=q^{\pm(J-I)/2}$ and $\lam=q^{(I+J)/2}$ we obtain
\begin{equation}\label{high6a}
S_{12}(q^{(J-I)/2})_{i,j}^{i',j'}=\delta_{i+j,i'+j'}\Phi_{q^2}(i|j';q^{-2I},q^{-2J}),
\end{equation}
\begin{equation}\label{high6b}
S_{12}(q^{(I-J)/2})_{i,j}^{i',j'}=\delta_{i+j,i'+j'}q^{2Ij-2Ji'}\Phi_{q^2}(j|i';q^{-2J},q^{-2I}),
\end{equation}
\begin{equation}\label{high6c}
S_{12}(q^{(I+J)/2})_{i,j}^{i',j'}=\delta_{i+j,i'+j'}\Phi_{q^2}(i|i+j;q^{-2I},q^{-2I-2J}).
\end{equation}

The reduction \eqref{high6a} was first noticed in \cite{Bor14} and then generalized to the higher rank case
in \cite{KMMO16}. The weights $I$ and $J$ can take complex values and play the role of spectral parameters.
This case corresponds to the Povolotsky model \cite{Povol13}.

Note that  $S_{12}^{t_1}$  is no longer invertible in (\ref{high6a}-\ref{high6b})
 and we can not define the dual
reflection equation. This  is similar to the TASEP model where we can still define
integrable boundary conditions for TASEP as a limit from the more general ASEP model \cite{Ragoucy2014}.

\nsection{Recurrence relations for $K$-matrices}

We are interested in finding a general  solution of the reflection equation
\eqref{intr2} with the $R$-matrix \eqref{Sdef} for arbitrary higher weights.

The reflection equation \eqref{intr2} takes the form
\begin{equation}
S_{I,J}(x/y)K_I(x)S_{J,I}(y/\bar{x})K_J(y)=K_J(y)S_{IJ}(x/\bar{y})K_I(x)S_{JI}(\bar{y}/\bar{x}).
\label{recur1}
\end{equation}
We are only interested in non-diagonal solutions of \eqref{recur1}, since any diagonal
$K$-matrix satisfying \eqref{recur4} below will be proportional to the identity matrix.

It is well known that for the case of $I=J=1$ the equation \eqref{recur1} admits a 4-parametric solution
of $2\times2$ matrix $K_1(x)$ \cite{vega1993}. In addition, compatibility conditions of \eqref{recur1} lead to the
following restriction
\beq
\bar{x}=1/x. \label{recur2}
\eeq

A general non-diagonal solution for $K_1(x)$ has the following form
\beq
K_1(x)_{i}^{i'}=
\left(
\begin{array}{cc}
\frac{\ds t_-}{\ds q\nu}-q\nu t_++x^2(t_--{\ds t_+})&\mu^{-1}t_-(x^{2}-x^{-2})\\
\mu t_+(x^{2}-x^{-2})&\frac{\ds t_-}{\ds q\nu}-q\nu t_++x^{-2}(t_--{\ds t_+})
\end{array}
\right)_{i+1,i'+1}\label{recur3}
\eeq
where $t_+,t_-,\mu,\nu\in\mathbb{C}$ are arbitrary complex parameters. This parametrization
of $K_1(x)$ naturally appears from solving equations for $K_J(x)$,  $J>1$ below.

A stochasticity condition for $K_1(x)$
\beq
\sum_{i}K_1(x)_{i}^{i'}=\>\mbox{independent of }\>i'\label{recur4}
\eeq
has two solutions $\mu=1$ and $\mu=-t_-/t_+$. It is easy to see that these two solutions
are equivalent up to a reparametrization of the remaining parameters $t_\pm,\nu$.
It is convenient to choose a solution
\beq
\mu=1. \label{recur5}
\eeq
Later on we will see that the $K$-matrix depends on the parameter $\mu$
in a simple way and one can set $\mu=1$ without loss of generality.

Let us notice that the stochastic $K$-matrix (2.25) with parameters $\alpha,\gamma$ from \cite{Ragoucy2014}
is obtained from \eqref{recur3} by a specialization
\beq
\mu=1,\quad t_+=\alpha,\quad t_-=\gamma,\quad
\frac{t_-}{q\nu}-q\nu t_++1+t_--t_+-q^2=0.\label{recur6}
\eeq
Now we substitute $I=1$ into the reflection equation \eqref{recur1}
and obtain
\beq
S_{1,J}(x/y)K_1(x)S_{J,1}(xy)K_J(y)=K_J(y)S_{1,J}(xy)K_1(x)S_{J,1}(x/y).\label{recur7}
\eeq
This is a linear system of recurrence relations for the matrix $K_J(y)$ with arbitrary $J$. Moreover,
we can keep $J$ as a complex parameter, since $L$-operators $S_{1,J}(x)$ and $S_{J,1}(x)$ are well defined
even for $J\in\mathbb{C}$.

In principle, its solution for integer $J$ is known and given by the fusion procedure \cite{Nepom92,Nepom07}.
However, we are interested in finding explicit formulas for matrix elements of $K_J(x)$
or their generating function.

Whether the solution $K_J(x)$ of \eqref{recur7} will satisfy
\eqref{recur1} with both $I,J\in\mathbb{C}$ is not clear. Most likely the answer is negative
because the equation \eqref{recur1} contains a double sum which is terminated either
by $I$ or $J$. If this double sum is infinite we can not use hypergeometric identities
similar to Sears' transformations. We have already seen this phenomenon with the crossing
unitarity relation.

We note that a situation with the Yang-Baxter equation  is different.
Due to the conservation law in \eqref{Rmatrix} internal sums in the Yang-Baxter
equation are terminated by external indices. This is the reason why
the solution \eqref{Rmatrix} can be analytically continued to complex $I$ and $J$
\cite{KMMO16}.

To find equations for $K_J(y)$ in \eqref{recur7}
we need to derive formulas for the $L$-operators $S_{1,J}(x)$ and $S_{J,1}(x)$.
Specifying $I=1$ and $J=1$ in \eqref{Rmatrix} and using \eqref{Sdef} one can obtain  after
straightforward calculations
\begin{eqnarray}
&&{S_{1,J}(x)_{i,j}^{i',j'}=
\left(
\begin{array}{cc}
\delta_{j,j'}\,q^j\frac{\ds [xq^{\frac{1+J}{2}-j}]}{\ds[xq^{\frac{1+J}{2}}]}&
\delta_{j,j'+1}\,xq^{j-\frac{J+1}{2}}\frac{\ds [q^{1+J-j}]}{\ds[xq^{\frac{1+J}{2}}]}\\\\
\delta_{j+1,j'}\,q^{j-\frac{J-1}{2}}\frac{\ds[q^{1+j}]}{\ds  x [xq^{\frac{1+J}{2}}]}&
\delta_{j,j'}\,q^{j-J}\frac{\ds [xq^{\frac{1-J}{2}+j}]}{\ds[xq^{\frac{1+J}{2}}]}
\end{array}
\right)_{i+1,i'+1}},\label{recur8}\\ \nonumber\\
&&{
S_{J,1}(x)_{j,i}^{j',i'}=
\left(
\begin{array}{cc}
\delta_{j,j'}\,q^{-j}\frac{\ds [xq^{\frac{1+J}{2}-j}]}{\ds[xq^{\frac{1+J}{2}}]}&
\delta_{j,j'+1}\,q^{\frac{J+1}{2}-j}\frac{\ds [q^{1+J-j}]}{\ds x[xq^{\frac{1+J}{2}}]}\\\\
\delta_{j+1,j'}\,x q^{\frac{J-1}{2}-j}\frac{\ds[q^{1+j}]}{\ds   [xq^{\frac{1+J}{2}}]}&
\delta_{j,j'}\,q^{J-j}\frac{\ds [xq^{\frac{1-J}{2}+j}]}{\ds[xq^{\frac{1+J}{2}}]}
\end{array}
\right)_{i+1,i'+1}},\label{recur9}
\end{eqnarray}
where all indices $i,j,i',j'\in\mathbb{Z}_+$ (or $\leq J$ for integer $J$) and we used a notation
\beq
[x]=x-x^{-1}. \label{recur10}
\eeq
Substituting (\ref{recur3}, \ref{recur8}-\ref{recur9}) into \eqref{recur7} we obtain a set of equations
polynomial in $x$. Decoupling with respect to $x$ we get 12  recurrence
relations for matrix elements $K_J(y)_j^l$. After some algebra one can see that only two of them
are linearly independent

\begin{equation}\label{recur11}\begin{aligned}
&\mu t_+q^{2+2J}(1-q^{2(j-J)})K_J(y)_j^{l+1}+\mu^{-1}t_-q^{2J}(1-q^{2+2l})K_J(y)_{j+1}^l+\\
&\nu^{-1}y^2{q^{2+J}}(q^{2j}-q^{2l})(t_--\nu^2q^2t_+) K_J(y)_{j+1}^{l+1}-\\
&\mu t_+y^4q^{2+2J}
(1-q^{2(1+l-J)})K_J(y)_{j+1}^{l+2}-\mu^{-1}t_-y^4q^{2J}(1-q^{4+2j})K_J(y)_{j+2}^{l+1}=0,
\end{aligned}
\end{equation}\\
\begin{equation}\label{recur12}\begin{aligned}
&\mu t_+q^{2(2+l+J)}(1-q^{2j-2J})K_J(y)_j^{l+1}+\mu^{-1}t_-q^{2(1+j+J)}(1-q^{2+2l})K_J(y)_{j+1}^l+\\
&{q^{2+2J}}(q^{2j}-q^{2l})(t_+-t_-)
K_J(y)_{j+1}^{l+1}-\\
&\mu t_+q^{2(1+j+J)}(1-q^{2(1+l-J)})K_J(y)_{j+1}^{l+2} -
\mu^{-1}t_-q^{2J+2l}(1-q^{4+2j})K_J(y)_{j+2}^{l+1}=0.
\end{aligned}
\end{equation}

A detailed analysis of these relations for arbitrary complex $J$ shows that any solution contains
two arbitrary parameters $K_J(y)_0^0$ and $K_J(y)_1^0$ and we can consistently choose
\beq
K_J(y)_{-k}^l=0,\quad k=1,2,\dots,\quad l=0,1,2\ldots\label{recur13}
\eeq
For any $J=1,2,3,\ldots$
we impose a terminating condition
\beq
K_J(y)_{J+1}^0=0. \label{recur14}
\eeq

The condition \eqref{recur14} determines $K_J(y)_1^0$ in terms of the  normalization
factor
$K_J(y)_0^0$. Once \eqref{recur14} is satisfied, a simple analysis of (\ref{recur11}-\ref{recur12}) shows
  that
\beq
K_J(y)_{J+1+j}^l=0, \quad \mbox{for}\quad j,l\geq0.\label{recur15}
\eeq
Let us notice that, in general, $K_J(y)_j^{J+1}\neq0$ for $0\leq j\leq J$, i.e. there is no termination
with respect to the index $l$. However, \eqref{recur15} already ensures that all sums
in \eqref{recur7} are finite for $J\in\mathbb{Z}_+$.

In particular, for $J=1$ we reproduce  a solution \eqref{recur3} and for $J=2$  explicit
formulas for the $K$-matrix are given in Appendix B.
For the $J=2$ untwisted  $R$-matrix \eqref{Rmatrix}   the corresponding $K$-matrix was first obtained
in \cite{Inami96}.

Now we will show that the condition \eqref{recur4} with $\mu=1$ is compatible with
(\ref{recur11}-\ref{recur12}) for any $J=1,2,3,\ldots$.

First, we introduce two quantities
\beq
S_l=\sum_{j=0}^J K_J(y)_j^l,\quad T_l=\sum_{j=0}^Jq^{2j} K_J(y)_j^l,\quad l\geq0.
\eeq
Summing up \eqref{recur12} over $j$ and  taking \eqref{recur13} and \eqref{recur15} into account
we can express $T_l$ in terms of $S_l$ and $S_{l\pm1}$. Summing up  \eqref{recur11}
over $j$ and substituting $T_l$ we observe  that a constant solution $S_l=S$ exists provided
that $\mu=1$ or $\mu=-t_-/t_+$ independent of $J$.

Indeed, we solved (\ref{recur11}-\ref{recur12}) for $J=1,2,3$ and checked that
up to an overall normalization the $K$-matrix is stochastic at $\mu=1$.

\nsection{Special solutions of the reflection equation}

In this section we first analyze lower- and upper- triangular solutions of the reflection equation.
Let us notice that the defining relations (\ref{recur11}-\ref{recur12}) become trivial
for diagonal $K$-matrices. So we assume that either $t_+$ or $t_-$ is not equal to $0$.

First we set $t_+=0$. Then it is easy to see that a solution to (\ref{recur11}-\ref{recur12}) has
an upper-triangular form and a simple analysis shows that
\beq
K_J(y)_j^l=\mu^{j-l}\Phi_{q^2}\left(j\Big{|}l;-\frac{y^2}{\nu q^J},-\frac{1}{y^2\nu q^J}\right).\label{spec1}
\eeq
The parameter $t_-$ becomes an overall  factor in (\ref{recur11}-\ref{recur12})
and can be set to $1$.
 The solution \eqref{spec1} is well defined even for $J\in\mathbb{C}$.
 When $J\in\mathbb{Z}_+$, both indices run the values $0\leq j,l\leq J$.
Due to
the property \eqref{sumrule},  the matrix \eqref{spec1} at $\mu=1$ is stochastic
\beq
\sum_{j=0}^\infty K_J(y)_j^l=1, \label{spec1a}
\eeq
where the sum terminates at $j=l$.

Similarly, if we set $t_-=0$, then the solution of (\ref{recur11}-\ref{recur12}) has a lower-triangular form
\beq
K_J(y)_j^l=c_J(\mu q^2)^{j-l}\frac{(q^2;q^2)_l}{(q^2;q^2)_j}\frac{(q^{-2J};q^2)_j}{(q^{-2J};q^2)_l}\>
\Phi_{q^2}\left(l{\Big{|}}j;-\frac{y^2\nu }{q^{J-2}},-\frac{\nu }{y^2q^{J-2}}\right).\label{spec2}
\eeq
where $c_J$ is the normalization factor.
All matrix elements in \eqref{spec2} become zero for $j>J$, $J\in\mathbb{Z}_+$ due to the factor
$(q^{-2J};q^2)_j$. If we choose
\beq
c_J=y^{4J}\frac{\ds\left(-\frac{\nu }{y^2q^{J-2}};q^2\right)_J}{\ds\left(-\frac{y^2\nu }{q^{J-2}};q^2\right)_J},
\label{spec2a}
\eeq
then we obtain $K_J(y)_J^J=1$ by using the definition \eqref{high3} of the $\Phi$ function.
Applying the $q$-Vandermonde summation formula \eqref{A5} it is easy to check that
\beq
\sum_{j=0}^J K_J(y)_j^l=1, \label{spec2b}
\eeq
i.e. the matrix \eqref{spec2} is also stochastic for any $J\in\mathbb{Z}_+$.

From (\ref{spec1}-\ref{spec2}) we can construct
upper- and lower-  triangular solutions of the dual reflection equation
using the mapping \eqref{spec4}.

One  can specify spectral parameters $x$ and $y$ in the reflection equation \eqref{recur1} such that
all $R$-matrices degenerate to a single $\Phi$ function as in (\ref{high6a}-\ref{high6c}).
This is achieved by setting
\beq
x=q^{I/2}, \quad y=q^{J/2}. \label{spec7}
\eeq
Under this specialization the $R$-matrices in  \eqref{recur1} degenerate to
different limits, i.e. \eqref{high6b}, \eqref{high6c} in the LHS and  \eqref{high6c}, \eqref{high6a}
in the RHS. In particular, the $R$-matrix $S_{J,I}(y/\bar x)$ degenerates into \eqref{high6c}
which is no longer invertible.

One can ask whether it is possible to start with the degenerate $R$-matrix \eqref{high6a} without difference
property
\begin{equation}\label{spec8}
S_{12}(x,y)_{i,j}^{i',j'}=\delta_{i+j,i'+j'}\Phi_{q^2}(i|j';x,y), \quad x=q^{-2I},
\quad y=q^{-2J},\quad x,y\in\mathbb{C}
\end{equation}
and construct solutions to the reflection equation
\begin{equation}
S_{12}(x,y)K_1(x,\bar x)S_{21}(y,\bar{x})K_2(y,\bar y)=K_2(y,\bar y)
S_{12}(x,\bar{y})K_1(x, \bar x)S_{21}(\bar{y},\bar{x}).
\label{spec9}
\eeq

We found that the equation \eqref{spec9} admits the following
upper-triangular solution
\beq
K(x,\bar x)_j^l=\Phi_{q^2}(j|l;zx,z\bar x) \label{spec10}
\eeq
where $z\in\mathbb{C}$ and parameters $\bar x$ and $\bar y$ in \eqref{spec9} are not constrained by \eqref{recur2}
and remain free. The reflection equation \eqref{spec9}
reduces to the 4th degree relation for $\Phi$ functions
\begin{align}
&\sum_{\beta_1,\beta_2}\Phi(\beta_1|\al;u,v)\Phi(\beta_2|\al+\beta-\beta_1;zx,zv)\Phi(\ga|\beta_1;x,u)
\Phi(\de|\beta_1+\beta_2-\ga;z y, zu)=\nonumber\\
&\sum_{\beta_1,\beta_2}\Phi(\ga|\beta_1;x,y)\Phi(\beta_1|\al;y,v)\Phi(\ga+\de-\beta_1|\bet_2;zx,zv)
\Phi(\beta_1+\beta_2-\al|\beta;zy,zu),\label{spec11}
\end{align}
where we dropped a subscript $q^2$ of the function $\Phi$,
 $\al,\bet,\ga,\de\in\mathbb{Z}_+$, $x,y,z,u,v\in\mathbb{C}$ and all summations
are finite and restricted by external indices. Surprisingly \eqref{spec11} is very hard
to prove. It reduces to some transformation of double generalized hypergeometric series
which we failed to identify.

Moreover, this identity can be directly generalized to
a higher rank $n>1$ by replacing the function $\Phi$ with its $U_q(A_n^{(1)})$ version from
\cite{KMMO16} with all indices replaced by their $n$-component analogs.
We checked a generalization of \eqref{spec11} for $n=1,2,3$ and external indices $\leq3$ and
leave it as a conjecture.

\nsection{A non-degenerate case}

In this section we study the general off-diagonal $K$-matrices when both parameters $t_\pm\neq0$.
First, we notice that any off-diagonal solution of (\ref{recur11}-\ref{recur12}) possesses a symmetry
\beq
K_J(y)_j^l=\left(-q^2\mu^2\frac{t_+}{t_-}\right)^{j-l}\frac{(q^{-2J};q^2)_j}{(q^{-2J};q^2)_l}
\frac{(q^2;q^2)_l}{(q^2;q^2)_j}\label{nondiag1}
K_J(y)_l^j.
\eeq
This can be established by substituting $K_J(y)_j^l$ from the LHS of \eqref{nondiag1} to
(\ref{recur11}-\ref{recur12}) and showing that the resulting recurrence relations
are equivalent to original ones.

Using this fact let us define
a new variable $t$ as
\beq
t^2=\frac{t_+}{t_-}\label{nondiag2}
\eeq
and
introduce  matrices $N_{j,l}$ by
\beq
K_J(y)_j^l=(-1)^lq^{2j}(\mu t)^{j-l} \frac{(q^2;q^2)_l}{(q^{-2J};q^2)_l}N_{j,l}.\label{nondiag3}
\eeq
It is easy to see from \eqref{nondiag1} that $N_{j,l}$ is symmetric
\beq
N_{j,l}=N_{l,j}. \label{nondiag4}
\eeq
Recursion relations (\ref{recur11}-\ref{recur12}) can be rewritten for $N_{j,l}$
\begin{align}
&q^{2j+2l}(1-q^{2(1+J-j)})N_{j-1,l}+q^{2(1+l+J)}(1-q^{2+2j})N_{j+1,l}+
q^{2(1+J+j)}(t^{-1}-t)N_{j,l}=\nonumber\\
&q^{2j+2l}(1-q^{2(1+J-l)})N_{j,l-1}+q^{2(1+j+J)}(1-q^{2+2l})N_{j,l+1}+
q^{2(1+J+l)}(t^{-1}-t)N_{j,l}\label{nondiag5}
\end{align}
\begin{align}
&y^{-2}q^{2j}(1-q^{2(1+J-j)})N_{j-1,l}+y^2q^{4+2J}(1-q^{2+2j})N_{j+1,l}+
q^{2+J+2j}(q^2t\nu -(t\nu)^{-1})N_{j,l}=\nonumber \\
&y^{-2}q^{2l}(1-q^{2(1+J-l)})N_{j,l-1}+y^2q^{4+2J}(1-q^{2+2l})N_{j,l+1}+
q^{2+J+2l}(q^2t\nu -(t\nu)^{-1})N_{j,l}.\label{nondiag6}
\end{align}
If we impose boundary conditions $N_{j,l}=0$ for any $j,l<0$, then a solution
to (\ref{nondiag5}-\ref{nondiag6}) will be symmetric in $j,l$ and depend
on two initial conditions, say $N_{0,0}$ and $N_{1,0}$.
Moreover, $N_{j,l}$ will depend only on three parameters, $y$, $t$ and $\nu$
and  we will omit this dependence from now on.

Recently very similar equations for higher rank $K$-matrices
were derived using a special coideal algebra of $U_q(A_{n}^{(1)})$ \cite{KOY18}.
The authors of \cite{KOY18} solved the analog of (\ref{nondiag5}-\ref{nondiag6})
using a matrix product  of local operators acting in
the auxiliary $q$-oscillator algebra. This approach is inspired by a 3D structure
of the $R$-matrix \eqref{Sfactor} and was developed by several authors
\cite{Bazhanov:2005as,Bazhanov:2008rd,KO12,Man14,KuP18}.

However, the equations for $K$-matrices in \cite{KOY18}
depend only on the spectral parameter  with no free parameters similar to $t$ and $\nu$
above. It would be very interesting to understand whether their approach
can be extended to find a matrix product solution of (\ref{nondiag5}-\ref{nondiag6}).
Unfortunately, we failed to do this and developed an alternative approach
using techniques coming from basic hypergeometric functions.

Let us introduce a generating function for matrix elements $N_{j,l}$
\beq
F(u,v)=\sum_{j,l=0}^\infty u^j v^l N_{j,l}\label{nondiag7}
\eeq
By \eqref{nondiag4} $F(u,v)$ is symmetric
\beq
F(u,v)=F(v,u).\label{nondiag8}
\eeq
From (\ref{nondiag5}-\ref{nondiag6}) we can derive a system of coupled $q$-difference equations
for $F(u,v)$

\beq
u(1-v/t)(1+vt)F(q^2u,v)-v(1-u/t)(1+ut)F(u,q^2v)-(u-v)(1+uvq^{-2J})F(q^2u,q^2v)=0,\label{nondiag9}
\eeq
\begin{align}
u\left(1+\frac{v}{q^{2+J}t\nu y^2}\right)\left(1-\frac{t\nu v}{q^J y^2}\right)&F(u,q^2 v)-
v\left(1+\frac{u}{q^{2+J}t\nu y^2}\right)\left(1-\frac{t\nu u}{q^J y^2}\right)F(q^2u,v)-\nonumber\\
(u-v)\left(1+\frac{uv}{q^2y^4}\right)&F(u,v)=0. \label{nondiag10}
\end{align}
When we derive equations for generating functions from recurrence relations, one can expect
extra boundary terms in difference equations corresponding to initial conditions in \eqref{nondiag7},
see for example equation (4.4) in \cite{GIM94}.
However, since recurrence
relations for $N_{j,l}$ are consistent with terminating conditions $N_{j,l}=0$ for $j<0$ or $l<0$,
no boundary terms appear in (\ref{nondiag9}-\ref{nondiag10}). Expanding (\ref{nondiag9}-\ref{nondiag10})
in series in $u$ and $v$ one can check that coefficients for any solution of the form \eqref{nondiag7}
will solve recurrence relations for $N_{j,l}$.

We also note that a special choice of parameters in the  $K$-matrix \eqref{recur3} ensures that
all coefficients in (\ref{nondiag9}-\ref{nondiag10}) factorize. This was the main reason
for using such a parametrization.

\nsection{Construction of the generating function $F(u,v)$}

Instead of solving the system (\ref{nondiag9}-\ref{nondiag10}) in two variables $u,v$ we can exclude
shifts in $v$ and derive a 2nd order difference equation in $u$ only. The result reads
\begin{align}
q^2\left(1-\frac{u}{q^2t}\right)\left(1+\frac{tu}{q^2}\right)\left(1+\frac{uv}{q^4y^4}\right)
&[F(\frac{u}{q^2},v)-F(u,v)]+\nonumber\\
\left(1-\frac{tu\,\nu}{q^Jy^2}\right)\left(1+\frac{u}{q^{2+J}t y^2\nu}\right)
\left(1+\frac{uv}{q^{2+2J}}\right)
&[F({u}{q^2},v)-F(u,v)]-\nonumber\\
\frac{u^2}{q^6y^2}(1-q^{-2J})\Big((1-q^{2-2J})uvy^{-2}-q^{3-J}v[qt\nu]
+q^4[y^2]-&q^2v[t]y^{-2}\Big)F(u,v)=0,\label{gen1}
\end{align}
where $[x]$ for $x\neq0$ is defined in \eqref{recur10}.

Our goal is to construct a solution to \eqref{gen1} which is a polynomial in $u,v$ of the degree
$J$ for $J\in\mathbb{Z}_+$. Difference equations similar to (\ref{nondiag9}-\ref{nondiag10}) and
\eqref{gen1} have been studied by several authors \cite{IR91,GM94,GM94a,GIM94}. Their general solution
is given in terms of very well-poised non-terminating $\phiW$ series. If $\phiW$ series terminates,
then due to Watson's transformation formula (III.18) in \cite{Gasper} it can be transformed into
a terminating balanced ${}_4\phi_3$ series. So for $J\in\mathbb{Z}_+$ one can expect the answer
in terms of ${}_4\phi_3$ series.

A realization of this program has several difficulties. First, the 2nd order difference equation
for $\phiW$ (see (2.1) in \cite{GM94}) has the same structure as \eqref{gen1} but with all
coefficients factorized. This is not the case for \eqref{gen1}. However, this can be repaired
in the following way. Let us assume that a solution to \eqref{gen1} has the form
\beq
F(u,v)=\Psi(u)\phiW(u),\label{gen2}
\eeq
where $\phiW(u)$ solves the 2nd order equation in one variable with other parameters
fixed
(see (2.1) in \cite{GM94}). We will not give this equation here
because its explicit form is not important for further discussion.
We also assume that
$\Psi(u)$ satisfies the recurrence relation
\beq
\frac{\Psi(q^2u)}{\Psi(u)}=\rho(u)\label{gen3}
\eeq
with $\rho(u)$ being a rational function. We aim at finding  $\rho(u)$ which has a structure
of a simple product of a ratio of linear factors.
Then the function $\Psi(u)$ can be expressed in terms of $q$-Pochhammer symbols.

Now we substitute \eqref{gen2} into \eqref{gen1} and use the equation for $\phiW$  to exclude
the term with $\phiW(q^2 u)$. It results in the relation
\beq
(A(u)\rho(u)+B(u))\phiW(u)+(C(u)\rho(u)\rho(u/q^2)+D(u))\phiW(u/q^2)=0,\label{gen4}
\eeq
where $A(u),B(u),C(u),D(u)$ are known factors. Since $\phiW$ can not satisfy the first order
difference equation, both terms in \eqref{gen4} should be identically zero.
Solving these two relations with respect to $\rho(u)$ and $\rho(u/q^2)$
we get two compatibility conditions for the function $\rho(u)$. Further analysis
shows that one can choose parameters of $\phiW$ series in such a way that $\rho(u)$ is completely
factorized and $\Psi(u)$ is given by a product of $q$-Pochhammer symbols.
In this way one can find two linearly independent solutions $F_\pm(u,v)$ which are symmetric
in $u,v$ and solve \eqref{gen1}.

The main difficulty of this general approach is that both solutions $F_\pm(u,v)$ are nonterminating
even for integer $J$. We can form a linear combination
\beq
F(u,v)=\sum_{\epsilon=\pm}A_\ep F_\ep(u,v)\label{gen5}
\eeq
and demand that $F(u,v)$ is a polynomial in $u$ of the degree $J$ for $J\in\mathbb{Z}_+$.
This terminating condition can be written in terms of ${}_3\phi_2$ series and is very complicated.
Substituting this back to \eqref{gen5} we should obtain the desired polynomial solution
but the level of technical difficulties is so extreme that we did not succeed in finding
it in any reasonable form.

At least we can learn from the above calculations that a polynomial solution symmetric in $u$ and $v$
maybe expressible in terms of terminating very well-poised $\phiW$ series, i.e.
terminating balanced ${}_4\phi_3$ series. This is indeed the case as we will see below.

Let us start with the simpler case $v=0$ and construct $F(u,0)$.
If we substitute $v=0$ into \eqref{gen1}, we get a difference equation for $F_0(u)=F(u,0)$
\begin{align}
q^2\Big(1-\frac{u}{q^2t}\Big)\Big(1+\frac{tu}{q^2}\Big)
&[F_0({u}/{q^2})-F_0(u)]+
\Big(1-\frac{tu\,\nu}{q^Jy^2}\Big)\Big(1+\frac{u}{q^{2+J}t y^2\nu}\Big)
[F_0({q^2}u)-F_0(u)]-\nonumber\\
&\frac{u^2}{q^2y^2}(1-q^{-2J})
(y^2-y^{-2})F_0(u)=0.\label{gen6}
\end{align}
In fact, it is easier to solve difference equations for coefficients $N_{j,0}$ themselves.
Choosing $l=0$ in (\ref{nondiag5}-\ref{nondiag6}) and using $N_{j,-1}=0$ we get
\begin{align}
(q^{-2-2J}-q^{-2j})&N_{j-1,0}+(q^{-2j}-q^{2})N_{j+1,0}+
(q^{-2j}-1)[t]N_{j,0}=(1-q^{2})N_{j,1},\nonumber\\
\frac{(q^{2(j-J-1)}-1)}{q^2y^4}&N_{j-1,0}+(1-q^{2+2j})N_{j+1,0}-
\frac{(1-q^{2j})[qt\nu]}{q^{1+J}y^2}N_{j,0}=(1-q^{2})N_{j,1},\label{gen7}
\end{align}
where $[x]$  defined in \eqref{recur10}. Excluding $N_{j,1}$ from \eqref{gen7}
we get a three-term recurrence relation for $N_{j,0}$
\beq
(q^{2j};q^2)_2N_{j+1,0}+
(1-q^{2j})([t]+[qt\nu]q^{2j-J-1}/y^2)N_{j,0}-(1-q^{2(j-J-1)})(1-q^{2j-2}/y^4)N_{j-1,0}=0.\label{gen8}
\eeq
This relation is similar to a recurrence relation for Al-Salam-Chihara polynomials \cite{ASC76} (see (14.8.4)
in \cite{KLS}) and admits a terminating solution with $N_{j,0}=0$ for
$j>J$ in terms of  ${}_2\phi_1$ series. Using a contiguous relation \eqref{cont1}
 it is not difficult to check that
\begin{align}
&{N_{j,0}}=N_J\,
q^{2(J+1)(J-j)}\,t^{J-j}\frac{(q^{-2J};q^2)_{J-j}}{(q^2;q^2)_{J-j}}
\frac{\ds\left(-\frac{\nu}{y^2}q^{2+2j-J};q^2\right)_{J-j}}{\ds\left(\frac{q^{2j}}{y^4};q^2\right)_{J-j}}
\nonumber\\
&\times{}_{2}\phi_{1} \left(\begin{matrix} \left. \begin{matrix} q^{-2(J-j)}, -\nu y^2 q^{2-J}
\\ {\ds-\frac{\nu}{y^2}q^{2+2j-J}}\end{matrix} \right|  q^2,\ds\frac{q^J}{\nu t^2y^2},\end{matrix} \right),
\label{gen9}
\end{align}
where $N_J$ is the normalization factor which we will fix later from the stochasticity condition \eqref{spec2b}.

The generating function $F_0(u)$
is given by
\beq
F_0(u)=\sum_{j=0}^J u^j N_{j,0}. \label{gen10}
\eeq
To calculate $F_0(u)$ we first apply  the transformation \eqref{A6}  to \eqref{gen9}.
The result reads
\begin{align}
&{N_{j,0}}=N_J\,
q^{2(J+1)(J-j)}\,t^{J-j}\frac{(q^{-2J};q^2)_{J-j}}{(q^2;q^2)_{J-j}}
\frac{\ds\left(\frac{q^{2J}}{y^4},-\frac{q^2}{t^2};q^2\right)_{\infty}}
{\ds\left(-\frac{\nu q^{2+J}}{y^2},\frac{q^J}{\nu t^2y^2};q^2\right)_{\infty}}
\nonumber\\
&\times{}_{2}\phi_{1} \left(\begin{matrix} \left. \begin{matrix}\ds\frac{q^{-J}y^2}{\nu t^2}, -\nu y^2 q^{2-J}
\\ {\ds-q^2/t^2}\end{matrix} \right|  q^2,\ds\frac{q^{2j}}{y^4}\end{matrix} \right).
\label{gen11}
\end{align}
Expanding ${}_2\phi_1$ into series in $k$  and substituting the result into
 \eqref{gen10} we can calculate the sum over $j$ using
\eqref{A4a}
\beq
\sum_{j=0}^J u^j q^{2(J+1)(J-j)}\,t^{J-j}\frac{(q^{-2J};q^2)_{J-j}}{(q^2;q^2)_{J-j}}
\left(\frac{q^{2j}}{y^4}\right)^k=
u^Jy^{-4k}(q^2t/u;q^2)_J\frac{(u/t;q^2)_k}{(q^{-2J}u/t;q^2)_k}.\label{gen12}
\eeq
As a result we get the following expression for $F_0(u)$
\beq
F_0(u)=N_J u^J(q^2t/u;q^2)_J\frac{\ds\left(\frac{q^{2J}}{y^4},-\frac{q^2}{t^2};q^2\right)_{\infty}}
{\ds\left(-\frac{\nu q^{2+J}}{y^2},\frac{q^J}{\nu t^2y^2};q^2\right)_{\infty}}
{}_3\phi_2\left(\begin{matrix} \left. \begin{matrix}\ds u/t, -\nu y^2 q^{2-J},\frac{q^{-J}y^2}{\nu t^2}
\\ {\ds q^{-2J}u/t,-q^2/t^2\>\>\>\>\>
{\phantom{I^{I^I}}}}\end{matrix} \right|  q^2,\ds\frac{1}{y^4}\end{matrix} \right).\label{gen13}
\eeq
Applying the transformation \eqref{A7} with $q\to  q^2$ and
\beq
a=\frac{q^{-J}y^2}{\nu t^2},\quad b=u/t,\quad c=-\nu y^2 q^{2-J},\quad
d=q^{-2J}u/t,\quad e=-q^2/t^2 \label{gen14}
\eeq
we bring $F_0(u)$ back to the polynomial in $u$
\beq
F_0(u)=N_J u^J(q^2t/u;q^2)_J
\frac{\ds\left(\frac{q^{-J}}{\nu t^2 y^2};q^2\right)_J}{(y^{-4};q^2)_J}\>
{}_3\phi_2\left(\begin{matrix} \left. \begin{matrix}\ds q^{-2J},\> -\frac{u q^{-2-J}}{\nu t y^2},\>
\frac{q^{-J}y^2}{\nu t^2}\>
\\ {\ds q^{-2J}u/t,\>\frac{q^{-J}}{\nu t^2 y^2}\>\>\>\>\>
{\phantom{I^{I^I}}}}\end{matrix} \right|  q^2,\ds-q^{2+J}\frac{\nu}{y^2}\end{matrix} \right).\label{gen15}
\eeq
Finally applying \eqref{A8} we obtain
\beq
F_0(u)=N_J u^J(q^2t/u;q^2)_J\>{}_3\phi_2\left(\begin{matrix} \left. \begin{matrix}{\ds q^{-2J},\> -\nu y^2 q^{2-J},\>
\frac{q^{-J}y^2}{\nu t^2}}\>
\\ {\ds\frac{q^{-2J}u}{t},\>q^{2-2J}y^4}\end{matrix} \right|  q^2,q^2\end{matrix} \right).\label{gen16}
\eeq
The purpose of these calculations is to show how we arrived at \eqref{gen16}. Using contiguous relations
for ${}_3\phi_2$ \cite{GIM94} one can show that \eqref{gen16} indeed satisfies \eqref{gen6}. However, it is almost
impossible to guess this formula from contiguous relations for ${}_3\phi_2$.

Having the result \eqref{gen16} one can try to generalize it to the full generating function $F(u,v)$.
We expect it to be a terminating balanced ${}_4\phi_3$ series symmetric in $u$ and $v$.
The only possible candidate which reduces to $F_0(u)$ at $v=0$ is
\beq
F(u,v)=\overbar{N}_J(uv)^J(q^2t/u;q^2)_J(q^2t/v;q^2)_J\>
{}_4\phi_3\left(\begin{matrix} \left. \begin{matrix}{\ds q^{-2J},\,-\frac{uv}{q^{2+2J}}, -\nu y^2 q^{2-J},\,
\frac{q^{-J}y^2}{\nu t^2}}
\\ {\ds{q^{-2J}u}/{t},\,{q^{-2J}v}/{t},\,q^{2-2J}y^{4^{\phantom{I}}}}\end{matrix} \right|  q^2,q^2\end{matrix}
\right) \label{gen17}
\eeq
with
\beq
N_J=(-t)^Jq^{J(J+1)}\overbar{N}_J. \label{gen17a}
\eeq
Remarkably this is the answer. It solves both difference equations (\ref{nondiag9}-\ref{nondiag10}) which are
equivalent to contiguous relations
 (\ref{A9}-\ref{A10}) from Appendix A.

We also need to calculate the normalization factor ${\overbar{N}}_J$.
At $\mu=1$
we have from \eqref{spec2b} and \eqref{nondiag3}
\begin{align}
&\sum_{j,l=0}^J(q^2t)^jv^lN_{j,l}=
\sum_{l=0}^J(-vt)^l\frac{(q^{-2J};q^2)_l}{(q^2;q^2)_l}\Big(\sum_{j=0}^JK_J(y)_j^l\Big)=\nonumber\\
&\sum_{l=0}^J(-vt)^l\frac{(q^{-2J};q^2)_l}{(q^2;q^2)_l}=(-vt q^{-2J};q^2)_J,\label{gen18}
\end{align}
where we used \eqref{A4a}. Therefore, a correct normalization of the generating function $F(u,v)$ is given by
\beq
F(q^2t,v)=(-vtq^{-2J};q^2)_J.\label{gen19}
\eeq

Now we note that a pre-factor in \eqref{gen17} has a zero at $u=q^2t$ and only the last term with $k=J$ in the series
expansion
of ${}_4\phi_3$ has a pole. Therefore, only this last term survives in the limit $u\to q^2t$
\beq
F(q^2t,v)=\overbar{N}_J(q^4tv)^J(q^2t/v)_J\frac{\ds\Big(q^{-2J},-vtq^{-2J}, -\nu y^2 q^{2-J},
\frac{q^{-J}y^2}{\nu t^2};q^2\Big)_J}{(q^2,{q^{-2J}v}/{t},\,q^{2-2J}y^{4};q^2)_J}
\lim_{\lambda\to1}\frac{(\lambda^{-1};q^2)_J}{(q^{2-2J}\lambda;q^2)_J}.\label{gen20}
\eeq
Comparing the result with \eqref{gen19}
we obtain
\beq
\overbar{N}_{J}={q^{-2J(J+1)}}\frac{y^{4J}}{t^{2J}}\frac
{\ds(y^{-4};q^2)_J}{\ds\Big(-\nu y^2 q^{2-J},
\frac{q^{-J}y^2}{\nu t^2};q^2\Big)_J}.\label{gen21}
\eeq
This is the correct stochastic normalization of the generating function \eqref{gen17}.

To calculate matrix elements $N_{j,l}$ we need to expand the generating function \eqref{gen17}
back into series in $u$ and $v$. We can do this using the identity
\beq
u^J\frac{(q^2t/u;q^2)_J}{(q^{-2J}u/t;q^2)_k}=
(-1)^J t^J q^{J(J+1)}\sum_{n=0}^{J-k}\frac{(q^{-2(J-k)};q^2)_n}{(q^2;q^2)_n}\Big(\frac{u}{t}\Big)^n.\label{calc1}
\eeq
Expanding ${}_4\phi_3$ into series in $k$, using (\ref{gen21}-\ref{calc1}) together with \eqref{A5}
and replacing $k\to J-k$, we can calculate a matrix element
$N_{j,l}$ as a double sum
\begin{align}
N_{j,l}=
\sum_{k=0}^J\sum_{s=0}^{\min(j,l)}
\frac{(-1)^sq^{(k-2s)(k+1)}}{t^{j+l-2(k+s)}}
\frac{\ds(q^{-2J},y^{-4};q^2)_k }
{\big(q^2,\mathLarge{-\frac{q^{-J}}{\nu y^2},\frac{q^{2-J}\nu t^2}{y^2}};q^2\big)_k}
\frac{(q^{-2(J-k)};q^2)_s}{(q^2;q^2)_s}\prod_{m=\{j,l\}}\frac{(q^{-2k};q^2)_{m-s}}{(q^2;q^2)_{m-s}}
.\label{calc2}
\end{align}

In Appendix B we give explicit formulas for $N_{j,l}$ with
$J=1,2$. The $K$-matrix is now obtained from \eqref{nondiag3}. It automatically satisfies the stochasticity
condition \eqref{spec2b}.

\nsection{Conclusion}

In this paper we considered the problem of finding a general explicit solution for the reflection equation
related to the higher spin representations of the stochastic six vertex model.
We found a set of recurrence relations for matrix elements of the $K$-matrix and solved them
explicitly in lower- and up- triangular cases.
In the general case we expressed the generating function for matrix elements in terms of the terminating
balanced ${}_4\phi_3$ series. By expanding it we obtained the expression for matrix elements of the $K$-matrix
in the form of a double sum \eqref{calc2}. It would be interesting to understand whether
this formula can be rewritten in the form
of a single sum, i.e. some basic hypergeometric function.

Here we did not address the problem of positivity of matrix elements of the $K$-matrix.
However, for the case $J=1$
it is well known that all elements of the $R$-matrix and $K$-matrix can be chosen in a positive regime.
Since the higher spin $R$- and $K$-matrices can be built with fusion from elementary ones, we expect
that positivity holds for any $J$.

Another interesting problem is a connection of our results with a 3D approach
\cite{Bazhanov:2005as,Bazhanov:2008rd,KO12,Man14,KuP18}. In \cite{KOY18}
a matrix product solution to the reflection equation associated with a certain
coideal subalgebra of $U_q(A_n^{(1)})$ was constructed. Their defining equations for $n=1$
are very similar to our (\ref{nondiag5}-\ref{nondiag6}) but do not have any free
parameters except the spectral one. It is an important question whether it is possible
to find a matrix product solution to (\ref{nondiag5}-\ref{nondiag6}) with arbitrary
$\nu$ and $t$  equivalent to \eqref{calc2}.
If the answer is positive,
then a generalization to higher ranks
should be possible\footnote{After submitting this work A. Kuniba informed
us that their defining equations of the $K$-matrix at $n=1$ in \cite{KOY18} allow a generalization
which is equivalent to (\ref{nondiag5}-\ref{nondiag6}) after a certain transformation.
However, a matrix product solution for this more general case is not known.}.
We plan to address these questions in the next publication.

\section*{Acknowledgments}

We would like to thank Vladimir Bazhanov, Ivan Corwin, Jan De Gier, Ole Warnaar and Michael Wheeler  for
their interest to this work and useful discussions. V.M. would also like to thank
Eric Rains for his advice on the system (\ref{nondiag9}-\ref{nondiag10})
during Rainsfest in Brisbane in  October 2018, Ole Warnaar for his advice on the identity \eqref{spec11}
and Atsuo Kuniba for sending their work \cite{KOY18} and comments on the system (\ref{nondiag5}-\ref{nondiag6}).
This work was supported by the Australian Research Council, grant DP180101040.

\appendixtitleon
\begin{appendices}
\numberwithin{equation}{section}

\section{}
Here we list standard definitions in $q$-series which we need in the main text
\begin{align}
(a;q)_{\infty} &:= \prod_{i=0}^{\infty} (1-aq^{i}),\\
(a;q)_{n} &:= \frac{(a;q)_{\infty}}{(aq^{n};q)_{\infty}},\label{A1}
\end{align}
\begin{align}
(a_{1},\dots,a_{m};q)_{n} = \prod_{i=1}^{m}(a_{i};q)_{n},\label{A2}
\end{align}
\begin{align}
{n \brack m}_{q} := \frac{(q;q)_{n}}{(q;q)_{n-m}(q;q)_{m}}.\label{A3}
\end{align}

We define a basic hypergeometric series ${}_{r+1}\phi_{r}$ by
\begin{align}\label{A4}
{}_{r+1}\phi_{r} \left(\begin{matrix} \left. \begin{matrix} a_{1}, a_{2},
\dots,a_{r+1} \\  \phantom{a_1,}b_{1}, \dots,  b_{r}\phantom{w} \end{matrix} \right|  q,x \\
\end{matrix} \right) = \sum_{i\geq0} \frac{(a_1,\ldots,a_{r+1};q)_{i}}{
(q,b_1,\ldots,b_{r};q)_{i}}\> x^{i}.
\end{align}
We also need several summation formulas and transformations of such series which we list below.
Before each transformation we give its number in \cite{Gasper}.

The $q$-binomial theorem (II.3)
\beq
{}_{1}\phi_{0}(a;-;q,z)=\frac{(az;q)_\infty}{(z;q)_\infty},\quad |z|<1,\label{A4a}
\eeq
the $q$-Vandermonde sum (II.6)
\beq
{}_{2}\phi_{1} \left(\begin{matrix} \left. \begin{matrix} q^{-n}, a
\\ c \end{matrix} \right|  q,q\end{matrix} \right)=a^n\frac{(c/a;q)_n}{(c;q)_n},\label{A5}
\eeq
Heine's transformation (III.2)
\beq
{}_{2}\phi_{1} \left(\begin{matrix} \left. \begin{matrix} a,\>b
\\ c \end{matrix} \right|  q,z\end{matrix} \right)=
\frac{(c/b,bz;q)_\infty}{(c,z;q)_\infty}
{}_{2}\phi_{1} \left(\begin{matrix} \left. \begin{matrix} abz/c,b
\\ bz \end{matrix} \right|  q,c/b\end{matrix} \right),\label{A6}
\eeq
transformations of ${}_3\phi_2$ series (III.9) and (III.13)
\beq
{}_{3}\phi_{2} \left(\begin{matrix} \left. \begin{matrix} a,\>b,\>c\>
\\ d,\>e \end{matrix} \right|  q,\ds\frac{de}{abc}\end{matrix} \right)=
\frac{(e/a,de/bc;q)_\infty}{(e,de/abc;q)_\infty}
{}_{3}\phi_{2} \left(\begin{matrix} \left. \begin{matrix} a,\>d/b,\>d/c\>
\\ d,\>de/bc \end{matrix} \right|  q,\ds\frac{e}{a}\end{matrix} \right),\label{A7}
\eeq
\beq
{}_{3}\phi_{2} \left(\begin{matrix} \left. \begin{matrix} q^{-n},\>b,\>c\>
\\ d,\>e \end{matrix} \right|  q,\ds\frac{deq^n}{bc}\end{matrix} \right)=
\frac{(e/c;q)_n}{(e;q)_n}
{}_{3}\phi_{2} \left(\begin{matrix} \left. \begin{matrix} q^{-n},\>c,\>d/b\>
\\ d,\>cq^{1-n}/e \end{matrix} \right|  q,q\end{matrix} \right),\label{A8}
\eeq
Sears's transformation (III.16) for terminating balanced ${}_{4}\phi_{3}$ series
\begin{align}\label{Sears}
{}_{4}\phi_{3} \left(\begin{matrix} \left. \begin{matrix} q^{-n},  a,b,c \phantom{I}\\
\phantom{q^{-n},}d, e,  f\phantom{I}\end{matrix} \right|  q,q \\ \end{matrix} \right) &=
\frac{\left(a,\ds\frac{ef}{ab},\frac{ef}{ac};q\right)_{n}}
{\left(e,f,\ds\frac{ef}{abc};q\right)_{n}}\>
{}_{4}\phi_{3} \left(\begin{matrix} \left. \begin{matrix} \ds q^{-n}, & \ds\frac{e}{a},
&\ds\frac{f}{a},&\ds\frac{ef}{{abc_{\phantom{I}}}}\\
&\ds\frac{ef}{ab},& \ds\frac{ef}{ac}, & \ds\frac{{q^{1-n}}^{\phantom{I}}}{a}\end{matrix}
\right|  q,q \\ \end{matrix} \right)
\end{align}
provided that $def=abcq^{1-n}$.

The function ${}_{2}\phi_{1}$ satisfies the following contiguous relation
\beq
z(1-a)(b-c)\,{}_{2}\phi_{1} (a_+,c_+)+
(1-c)(q-c)\,{}_{2}\phi_{1} (a_-,c_-)+
(1-c)(c-q+(a-b)z)\,{}_{2}\phi_{1} =0,\label{cont1}
\eeq
where we used the standard notation $a_\pm=aq^{\pm1}$, etc and
dropped arguments of the ${}_{2}\phi_{1}$ function which do not change.
This is a direct consequence of Heine's contiguous relations (p.425 in \cite{NIST}).

The terminating balanced ${}_4\phi_3$ series defined in the LHS of \eqref{Sears}
satisfies
\beq
(1-a)(d-e)\,{}_4\phi_3(a_+,d_+,e_+)-(1-d)(a-e)\,{}_4\phi_3(e_+)+(1-e)(a-d)\,{}_4\phi_3(d_+)=0,\label{A9}
\eeq
\begin{align}
&e(1-e)(b-d)(c-d)(1-dq^n){}_4\phi_3(d_+)-
d(1-d)(b-e)(c-e)(1-eq^n){}_4\phi_3(e_+)\nonumber\\
&+(d-e)(1-d)(1-e)(bc-deq^n){}_4\phi_3(a_-)=0.\label{A10}
\end{align}
Relations (\ref{A9}-\ref{A10}) can be proved by specializing contiguous relations for very well-poised
$\phiW$ series \cite{IR91} to the terminating case.

\section{}

In this appendix we will give explicit formulas for the matrix $N_{j,l}$, $0\leq j\leq l\leq J$ for $J=1,2$.

Since the matrix $N_{j,l}$ is symmetric, we give only the upper-triangular elements. The normalization
is chosen in such a way that the $K$-matrix given by \eqref{nondiag3} satisfies stochasticity condition
\eqref{spec2b}
 at $\mu=1$.
 For $J=1$
\begin{align}
&N_{0,0}=\frac{y^2(1+\nu q y^2-\nu q t^2(\nu q+y^2))}
{(y^2-\nu q t^2)(1+\nu q y^2)},\nonumber\\
&N_{0,1}=-\frac{\nu t(1-y^4)}{q(y^2-\nu q t^2)(1+\nu q y^2)},\nonumber\\
&N_{1,1}=\frac{\nu q+ y^2-\nu q t^2(1+\nu q y^2)}
{q^4(y^2-\nu q t^2)(1+\nu q y^2)}\label{D1}
\end{align}
and for $J=2$
\begin{align}
&N_{0,0}=\frac{\nu t^2(y^2+\nu q^2)[\nu q^2 t^2(\nu+y^2)-(1+q^2)(1+\nu y^2)]+
(1+\nu y^2)(1+\nu q^2 y^2)}{(\nu t^2/y^2,-\nu y^2;q^2)_2},\nonumber\\
&N_{0,1}=\frac{\nu t}{q^2y^2}\frac{(1-y^4)(1+q^2)\left(\nu t^2(y^2+\nu q^2)-1-\nu y^2\right)}
{(\nu t^2/y^2,-\nu y^2;q^2)_2},\nonumber\\
&N_{0,2}=\frac{\nu^2 t^2}{q^4y^4}\frac{(1-y^4)(q^2-y^4)}{(\nu t^2/y^2,-\nu y^2;q^2)_2},\nonumber\\
&N_{1,1}=
\frac{1+q^2}{q^6}+\frac{\nu(1+q^2)(1-y^4)(q^2y^2(1+\nu y^2)+t^2(\nu q^2+y^2)(1-\nu q^2 y^2)-\nu q^2 t^4(1+\nu y^2))}
{{q^6y^4(\nu t^2/y^2,-\nu y^2;q^2)_2}},\nonumber\\
&N_{1,2}=\frac{\nu t}{q^8y^4}\frac{(1+q^2)(1-y^4)\left(\nu q^2t^2(1+\nu y^2)-y^2-\nu q^2\right)}
{(\nu t^2/y^2,-\nu y^2;q^2)_2},\nonumber\\
&N_{2,2}=\frac{(\nu+y^2)(\nu q^2+y^2)-\nu t^2(1+q^2)(\nu q^2+y^2)(1+\nu y^2)+\nu^2q^2 t^4(1+\nu y^2)(1+\nu q^2y^2)}
{q^{10}y^4(\nu t^2/y^2,-\nu y^2;q^2)_2}.\label{D2}
\end{align}

\end{appendices}

\newcommand\oneletter[1]{#1}
\providecommand{\href}[2]{#2}\begingroup\raggedright

\bibliographystyle{utphys}

\end{document}